\documentclass{nature}

\pdfoutput=1

\usepackage{graphicx}
\usepackage{color}
\usepackage{hyperref}
\usepackage{longtable}

\usepackage{lineno}

\newcommand{\msun}{${\rm M_{\sun}}$}

\def\ltsima{$\; \buildrel < \over \sim \;$}
\def\simlt{\lower.5ex\hbox{\ltsima}}
\def\gtsima{$\; \buildrel > \over \sim \;$}
\def\simgt{\lower.5ex\hbox{\gtsima}}
\def\kms{{\rm\,km\,s^{-1}}}
\def\mas{{\rm\,mas}}
\def\masyr{{\rm\,mas/yr}}
\def\kpc{{\rm\,kpc}}

\def\msun{{\rm\,M_\odot}}

\def\pc{{\rm\,pc}}

\newcommand{\fmmm}[1]{\mbox{$#1$}}

\newcommand{\scnp}{\mbox{\fmmm{''}}}


\def\deg{^\circ}
\def\degg{\hbox{$\null^\circ$\hskip-3pt .}}

\def\s{\ifmmode \widetilde \else \~\fi}
\def\={\overline}

\def\spose#1{\hbox to 0pt{#1\hss}}

\def\lta{\mathrel{\spose{\lower 3pt\hbox{$\mathchar"218$}}
     \raise 2.0pt\hbox{$\mathchar"13C$}}}
\def\gta{\mathrel{\spose{\lower 3pt\hbox{$\mathchar"218$}}
     \raise 2.0pt\hbox{$\mathchar"13E$}}}
\def\Dt{\spose{\raise 1.5ex\hbox{\hskip3pt$\mathchar"201$}}}    
\def\dt{\spose{\raise 1.0ex\hbox{\hskip2pt$\mathchar"201$}}}    

\def\dotsfill{\leaders\hbox to 1em{\hss.\hss}\hfill}

\def\Gyr{{\rm\,Gyr}}

\def\ltsima{$\; \buildrel < \over \sim \;$}
\def\gtsima{$\; \buildrel > \over \sim \;$}
\def\lsim{\lower.5ex\hbox{\ltsima}}
\def\gsim{\lower.5ex\hbox{\gtsima}}

\title{Identification of the Long Stellar Stream of the \\Prototypical Massive Globular Cluster $\omega$ Centauri}

\author{ Rodrigo A. Ibata$^{1}$, Michele Bellazzini$^{2}$, Khyati Malhan$^{1,3}$, Nicolas Martin$^{1,4}$ and Paolo Bianchini$^{1}$}

\begin{document}

\maketitle

\begin{affiliations}
\item Observatoire Astronomique, Universit\'e de Strasbourg, CNRS, 11, rue de l'Universit\'e, F-67000 Strasbourg, France.
\item INAF - Osservatorio di Astrofisica e Scienza dello Spazio di Bologna, Via Gobetti 93/3, 40129 Bologna, Italy
\item The Oskar Klein Centre for Cosmoparticle Physics, Department of Physics, Stockholm University, AlbaNova, 10691 Stockholm, Sweden
\item Max-Planck-Institut f\"ur Astronomie, K\"onigstuhl 17, D-69117, Heidelberg, Germany
\end{affiliations}

\newpage	

\begin{abstract}
Omega Centauri ($\omega$ Cen) is the most massive globular cluster\cite{Harris:1996fr} of the Milky Way, and possesses many peculiar properties. In particular, the cluster contains distinct multiple stellar populations, with a large spread in metallicity\cite{2010ApJ...722.1373J} and different kinematics as a function of light elements abundance\cite{2018ApJ...853...86B}, implying that it formed over an extended period of time\cite{2012ApJ...746...14M}. This has lead to the suggestion that $\omega$ Cen is the remnant core of an accreted dwarf galaxy\cite{1988IAUS..126..603Z,2000LIACo..35..619M}. If this scenario is correct, $\omega$ Cen should be tidally limited, and one should expect to find tidal debris spread along its orbit\cite{2003MNRAS.346L..11B,2003ApJ...589L..89M,2004ApJ...616L.107I,2004MNRAS.350.1141T}. Here we use N-body simulations to show that the recently-discovered ``Fimbulthul'' structure\cite{2019arXiv190107566I}, identified in the second data release (DR2) of the Gaia mission\cite{2018A&A...616A...1G,2018A&A...616A...2L}, is the long sought-for tidal stream of $\omega$ Cen, extending up to $28\deg$ from the cluster. Follow-up high-resolution spectroscopy of 5 stars in the stream show that they are closely-grouped in velocity, and have metallicities consistent with having originated in that cluster. Guided by our N-body simulations, we devise a selection filter that we apply to Gaia data to also uncover the portion of the stream in the highly-contaminated and crowded field within $10\deg$ of $\omega$ Cen. Further modelling of the stream may help to constrain the dynamical history of the dwarf galaxy progenitor of this disrupting system and guide future searches for its remnant stars in the Milky Way.
\end{abstract}

Stellar tidal streams represent the remnants of ancient accretion events into a galaxy that have remained spatially and kinematically coherent\cite{Johnston:1996gv}, and are thus extremely important as tracers of the galaxy formation process. They are generally long-lived features, as long as they do not stray too close to the dense inner regions of the host galaxy where dynamical times are short and the related precession and phase-mixing very fast. It is now realised that these structures are extremely powerful astrophysical tools, as they are sensitive probes of the acceleration field they inhabit\cite{Johnston:1999ix}, allowing us to study the underlying distribution of dark matter\cite{Ibata:2001be,2010ApJ...714..229L,2015ApJ...803...80K}, including the granularity or clumping of these dark materials\cite{2002MNRAS.332..915I,2002ApJ...570..656J,2012ApJ...748...20C,2016MNRAS.463..102E}.

The challenge now is to find a large sample of tidal streams that can be mapped with high-quality data, and measure their structural and kinematic properties. The recently-released Second Data Release (DR2) of the Gaia space mission provides an unparalleled new dataset to attempt such an endeavour. With 5-parameter astrometric solutions of $\approx$ 1.3 billion stars, Gaia DR2 gives excellent proper motion measurements of stars, as well as parallax measurements. 

We have developed a very pragmatic, data-driven approach to the problem of searching for stream-like structures in datasets such as Gaia DR2. We search for features that are spatially and kinematically coherent, possessing a distribution on the sky that resembles an orbit, together with distance properties and proper motion properties that are consistent with that orbit. Our algorithm\cite{2018MNRAS.477.4063M,2019arXiv190107566I}, which we call the {\tt STREAMFINDER}, works by examining every star in the Gaia survey in turn, sampling the possible orbits consistent with the observed photometry and kinematics, and finding the maximum-likelihood stream solution given a contamination model and a stream model. The adopted contamination model is an empirical model of the Milky Way in sky position, colour-magnitude and proper motion space (i.e. six-dimensions) constructed by smoothing the Gaia counts (here, on a $2\deg$ spatial scale), and building a Gaussian Mixture Model representation of the resulting smoothed Gaia data.

The adopted stream model is simple: we integrate to a given length along the sampled orbits (here $20\deg$), and over this length the stream counts are assumed to have uniform probability. Perpendicular to the stream model, the properties of all observables are taken to be Gaussian, so the model has a Gaussian physical width, a Gaussian dispersion in both proper motion directions, and a dispersion in distance modulus. The width and velocity dispersion parameters are chosen to be similar to the properties of known globular cluster streams. The algorithm returns the best-fit orbit for a given data point, the number of stars in the corresponding putative stream and the likelihood value relative to the model where the stream fraction is zero. By choosing those stars with stream solutions above a high significance level ($8\sigma$ in the present case), one can immediately derive a map of stream structures.

Because the Gaia parallaxes are relatively poor at constraining the distances to the survey stars, we use stellar populations models to convert the excellent Gaia photometry into trial line of sight distance values. To this end, we adopted the PARSEC isochrone models\cite{2012MNRAS.427..127B}, and our resulting maps are valid for a particular choice of population age and metallicity. To date, the {\tt STREAMFINDER} has allowed us to discover 14 new streams\cite{2018MNRAS.481.3442M,2018ApJ...865...85I,2019arXiv190107566I}, although we have so far assumed only a relatively narrow range of template stellar populations properties, due to the extremely high computational cost of the algorithm.

One of the structures we have identified in our stream map derived using a stellar population template of metallicity ${\rm [Fe/H]=-1.6}$ and age $12.5\Gyr$ is the ``Fimbulthul'' stream\cite{2019arXiv190107566I}, an $18\deg$-long feature, located at approximately $b=35\deg$ and following an arc between $\ell=-52\deg$ to $\ell=-32\deg$ (Figure~1a, red points). As the stream approaches $b=30\deg$ its contrast over the Galactic contamination drops precipitously (gray region in Figure~1a), and it becomes undetected by our algorithm. We display the proper motion and distance properties of these 309 stars in Figure~1b-d (red points), and list them in Table~1. Note that the stars still bound to $\omega$ Cen itself did not contribute in any way to the identification of this stream, as prior to running the detection software, we excised all stars from the input Gaia DR2 catalogue that lie within two tidal radii of all known globular clusters\cite{Harris:1996fr} (i.e., the white region within $1\degg6$ of $\omega$ Cen in Figure~4b).

Fortunately, two of the candidate stream members identified by the {\tt STREAMFINDER} (Figure~1e, orange points) have their radial velocities measured by the Gaia Radial Velocity Spectrometer (RVS; these measurements are only available for the very brightest stars with $G\simlt12.5$ in Gaia DR2). We fitted the sky-position, parallax, proper motion and radial velocity data of this sample of 309 stars using an orbit model (dotted line). Such a model is too simplistic if the progenitor of the stream is massive, because it does not take into account the large energy required to escape from a massive cluster (hence the escaping stars do not have the same orbital properties as their progenitor system). However, this nevertheless gives a first approximation of the orbital path and orbital properties of the stream. With the adopted Galactic mass model, the orbit fitted with a Markov Chain Monte Carlo procedure\cite{2018ApJ...865...85I} has a pericenter distance of $1.50\pm0.01\kpc$, an apocenter distance of $6.53\pm0.01\kpc$ and has highly retrograde motion with $L_z=490\pm1\kms \kpc$. We immediately noticed that these orbital properties are close to those recently-deduced\cite{2018A&A...616A..12G} for the massive globular cluster $\omega$ Cen (continuous line orbit in Figure~1) from Gaia DR2 data. 

We were able to obtain spectroscopic observations (see Methods) of six candidate member stars of the Fimbulthul stream with the ESPaDOnS high-resolution spectrograph on the Canada-France-Hawaii Telescope (CFHT). While one exposure was of insufficient signal-to-noise to yield a radial velocity measurement, all five others are found to be tightly grouped in radial velocity, confirming the reality of the Fimbulthul discovery. Four of the spectra are of sufficient quality to also allow metallicity measurements from the Ca~II infrared triplet, and yield values between ${\rm [Fe/H]= -1.36\pm0.11}$ to $-1.80\pm0.11$, consistent with an intrinsic metallicity dispersion, as would be expected for members of a population that was tidally disrupted from $\omega$ Cen. The results of these follow-up observations are summarised in Table~2.

We ran a series of N-body simulations of the tidal disruption of a system similar to $\omega$ Cen (see Methods), trying to reproduce the observed structure. The cluster progenitor was modelled as an axisymmetric system including internal rotation\cite{2013ApJ...772...67B}. There is very little freedom in the choice of orbit, given that even the line of sight distance to the cluster has recently been very well measured with RR~Lyrae variable stars\cite{2018AJ....155..137B}. Figure~2 shows the end-point of the  best simulation in the set. The blue points in Figure~2b are the simulation particles that are spatially coincident with the Fimbulthul stars (also shown in blue in Figure~1).

The measured proper motions of the Fimbulthul stars along the right ascension direction $\mu^*_\alpha (\equiv \mu_\alpha \cos(\delta)$, but henceforth, for convenience, we will drop the asterix superscript) and along the declination direction $\mu_\delta$, as a function of Galactic longitude $\ell$ match closely those of the N-body model (Figure 1b and 1c, respectively). The heliocentric radial velocities of the two stars measured by the Gaia RVS and with CFHT/ESPaDOnS also match the N-body model well (Figure 1e). The correspondence between this N-body model of $\omega$ Cen and the detected stream provides strong evidence that the Fimbulthul stream is the long-searched for (trailing) tidal arm of $\omega$ Cen. As we show in Figure~3a, the stellar population in the Fimbulthul stream is very similar to that of $\omega$ Cen, which provides further evidence for the link between these two structures. 

In our initial N-body explorations we modelled the cluster progenitor with a simple (non-rotating) King model, which cannot account for the complex internal dynamics of $\omega$ Cen\cite{2018ApJ...853...86B}, which is rotating and significantly flattened. The resulting N-body stream was substantially wider than the observed Fimbulthul stream; this shows that it is important to model the internal dynamics of the progenitor correctly, but also that observations of the structure and kinematics of the stream may allow us to deduce the past internal properties of the progenitor.

Our aim with these simulations is to illustrate the discovery of the stream and its connection to $\omega$ Cen. The full modelling of the structure in such a way as to reproduce faithfully the observations is beyond the scope of this contribution. Such modelling is very challenging because the precise morphology of a stream depends sensitively on the Galactic potential, and on the dynamical properties of the progenitor, including complex details such as its original shape and rotation properties. It may also require modelling the effects of time-dependent substructure in the Galaxy, such as the bar\cite{2015A&A...583A..76F} and spiral arms, as well as the effect of dynamical friction. We defer trying to obtain a good quantitative fit to the Fimbulthul stream to a future study.

We noticed that the kinematic behaviour of the particles in the N-body simulation that have just begun to leave $\omega$ Cen is very simple: they share approximately the same $\mu_\delta$ as the cluster, and display an approximately linear gradient in $\mu_\alpha$ as a function of $b$ (decreasing by $\approx 0.125\masyr$ for every degree in Galactic latitude). Using this prediction, we selected stars from the Gaia DR2 catalogue with proper motion measurements within $1\masyr$ (i.e. $\approx 26\kms$) of these criteria (we additionally required the parallax uncertainty to be less than $1\mas$, and for the parallax measurements to be consistent within $1\sigma$ with a distance in the range $[4,6]\kpc$). Stars obeying this selection function are shown in blue in Figure~3, where we compare the colour-magnitude distribution (CMD) of stars within $0.5\deg$ of the cluster centre (a), with a nearby reference field (b). It can be seen that this selection filter has not biassed us against selecting $\omega$ Cen stars. The difference between these two distributions motivated our choice to define the (green) CMD selection polygons which are intended to filter out contaminants. The resulting map of stars that have both the kinematic properties of the expected stream near $\omega$ Cen and that pass the colour and magnitude selection criterion are shown in Figure~4a. A clear narrow stream can be seen emanating from $\omega$ Cen, within $\pm 10\deg$ of the cluster, approximately along the path expected from the N-body model (Figure 2b). For comparison, Figure 4b additionally marks the positions of the Fimbulthul stream stars (red) and of the best N-body simulation (black). The knee in the N-body stream at $\ell \approx -57\deg$, $b\approx 20\deg$ appears to be mirrored in the Gaia DR2 map, pointing towards the Fimbulthul stream, as expected from the N-body simulation.

Although several surveys have detected clumps of stars at numerous locations around the sky \cite{2002ApJ...574L..39G,Meza:2005hn,Altmann:2005jc,2012ApJ...747L..37M,2017A&A...598A..58H,2018ApJ...860L..11K,2018MNRAS.478.5449M} with angular momentum properties that would be expected from the tidal debris of the progenitor of $\omega$ Cen, this is the first discovery of the actual tidal arms of the globular cluster. This confirms that $\omega$ Cen is the most massive tidally-disrupting globular cluster in our Galaxy. The N-body models we have run suggest that the system has been disrupting for at least $\approx 5\Gyr$, as shorter disruption times did not produce a feature similar to Fimbulthul. The models also suggest that there should be a substantial amount of debris elsewhere in the Galaxy (Figure 2a), which is mostly located at low Galactic latitude where it may be challenging to detect.

In future work it will be interesting to build on the findings reported here to devise a disruption model that can fully reproduce the tidal arms we have uncovered while simultaneously fitting accurately the remnant cluster. We expect that this may also require a more accurate model of the Galactic acceleration field (which this stream may contribute to constrain). With those additional constraints it will be interesting to examine the possible progenitors of $\omega$ Cen, and determine whether it may have been accreted onto the Milky Way as part of a more massive system. The reasonably good correspondence of our N-body model with the observed stream structures suggests that what we have identified here is material stripped from the cluster, and probably not stars of the hypothetical parent satellite galaxy that $\omega$ Cen was accreted with. Those stars, if they exist, may be searched for elsewhere probably at larger Galactocentric radii, where they would have been deposited during their catastrophic infall into our Galaxy.

\newpage

\begin{figure}
\includegraphics[width=15cm]{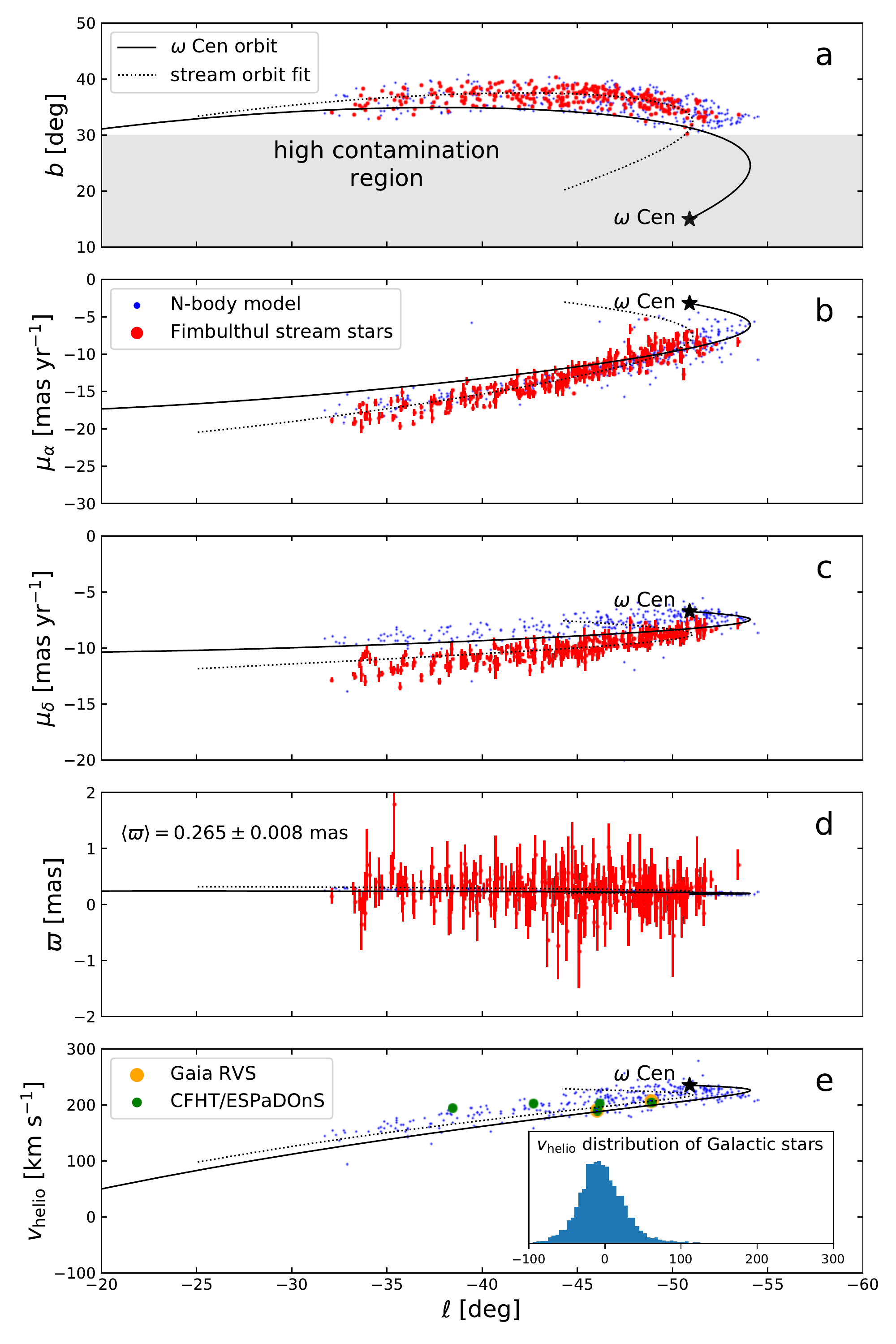}
\caption{Physical properties of the Fimbulthul stream as a function of Galactic longitude ($\ell$). The Gaia DR2 stars identified by the {\tt STREAMFINDER} algorithm are shown in red in all panels ($1\sigma$ error bars are shown), while the particles of the N-body simulation in the same region of sky are shown in blue. {\bf a} The sky distribution of the stars lies close to the backward-integrated orbit of $\omega$ Cen (continuous line). {\bf b} Proper motion $\mu_\alpha$ along the right-ascension ($\alpha$) direction. {\bf c} Proper motion $\mu_\delta$ along the declination ($\delta$) direction. (The measured proper motion\cite{2018A&A...616A..12G} of $\omega$ Cen is $\mu_\alpha=-3.1925\pm 0.0022 \masyr$, $\mu_\delta=-6.7445\pm 0.0019 \masyr$). {\bf d} Although the Gaia parallaxes ($\varpi$) appear not to be very informative, the absence of stars with parallaxes strongly inconsistent with the model shows that the sample is not obviously dominated by nearby contaminants. The weighted mean parallax is $\langle \varpi \rangle=0.265\pm0.008\mas$, although we caution the reader that this estimate does not account for any contamination that may be present in the sample. {\bf e} The heliocentric line of sight velocity $v_{\rm helio}$ from Gaia RVS (orange points) are shown, as well as the velocities from CFHT/ESPaDOnS (green points). The inset histogram shows the velocity distribution of the 5631 Gaia RVS stars with colours in the range of the Fimbulthul sample listed in Table~1 ($0.6 < G_{\rm BP}-G_{\rm RP} < 1.2$~mag) that are present in the region of sky where the Fimbulthul stream is found in panel (a). Clearly stars with radial velocities similar to those measured in the Fimbulthul stream are very rare. In all panels the dotted line corresponds to a simple orbit model fit to the stream. All stars shown here are identified by the algorithm as being $>8\sigma$ stream candidates. This means that for each of these stars, the algorithm finds a stream model that passes through the location of the star, with proper motion and distance consistent with that of the star, that stands out at $>8\sigma$ confidence above a smooth model of the Galaxy. Note that this does not mean that the individual stars are stream members with $8\sigma$ confidence.}
\end{figure}

\begin{figure}
\includegraphics[width=\hsize]{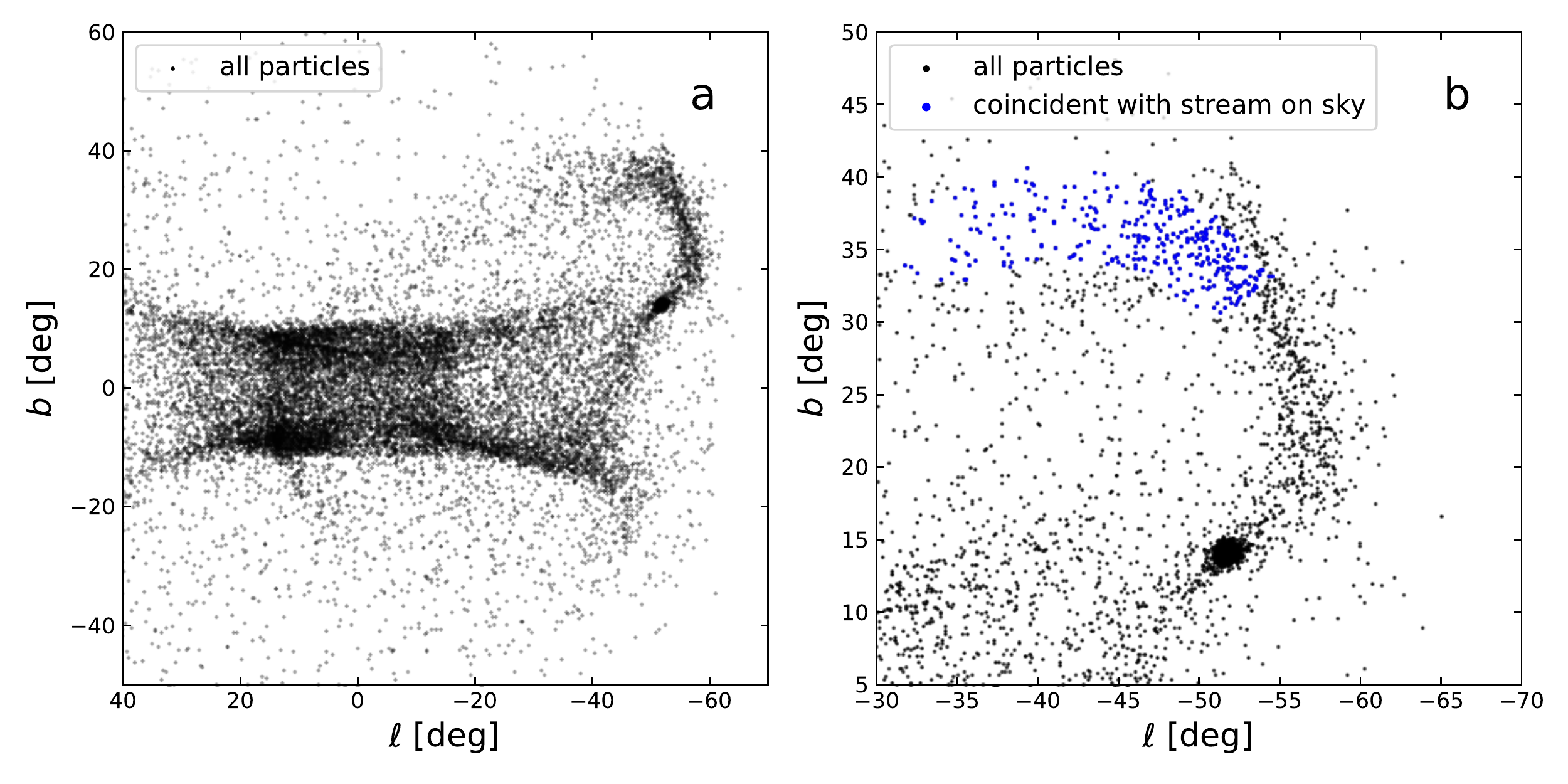}
\caption{Present day configuration of an N-body simulation of the disruption of a massive globular cluster on the orbit of $\omega$ Cen. {\bf a} The full simulation projected on Galactic coordinates. {\bf b} Zoom-in of the region of sky surrounding the cluster remnant. The blue particles were selected within a hand-drawn polygon to encompass the sky region containing the Fimbulthul stream stars in Figure~1a.}
\end{figure}

\begin{figure}
\includegraphics[width=\hsize]{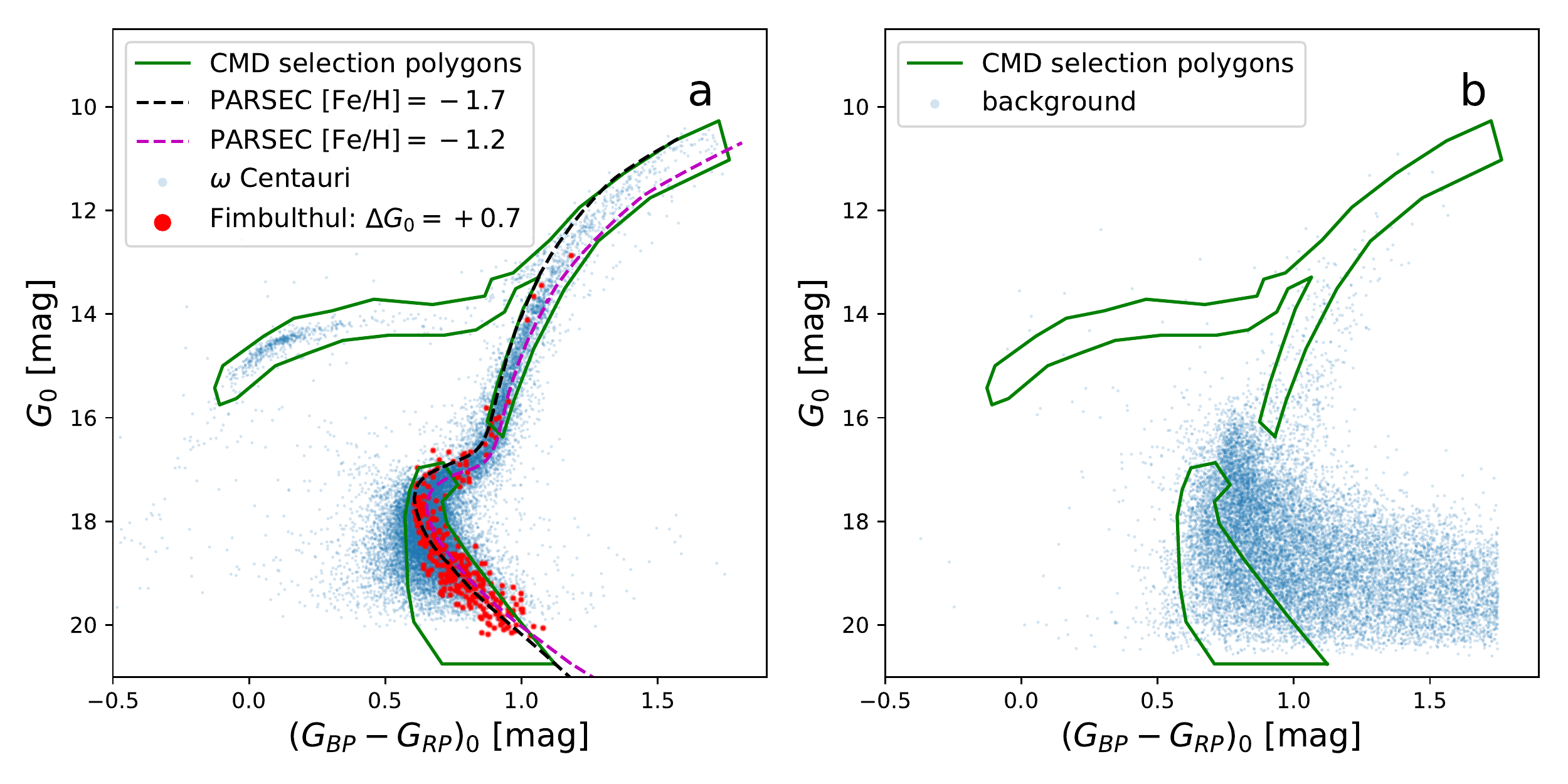}
\caption{Comparison of the colour-magnitude diagram of $\omega$ Cen and Fimbulthul in the Gaia photometric system. {\bf a} The stars within $0.5\deg$ of the cluster centre region and with kinematics consistent with cluster membership are shown in blue. The Fimbulthul stars identified by the {\tt STREAMFINDER} are displayed in red. These have been offset by 0.7~magnitudes to make the two distributions match (implying that the Fimbulthul sample is $\approx 1.5\kpc$ closer than $\omega$ Cen). PARSEC model isochones of age $12.5\Gyr$ and metallicity $-1.7$ and $-1.2$ are shown for comparison, these ${\rm [Fe/H]}$ values are representative of the spectroscopically measured range of the cluster \cite{Gratton:2011ie}. {\bf b} The colour-magnitude distribution of a nearby background field (covering the sky area $-70\deg < \ell <-40\deg$, $10\deg<b<50\deg$, but without those sources within a $10\deg$ radius circle of $\omega$ Cen). The green polygons in both panels show the two colour-magnitude selections chosen to select stars with properties similar to $\omega$ Cen, while minimising the number of contaminants.}
\end{figure}

\begin{figure}
\includegraphics[width=\hsize]{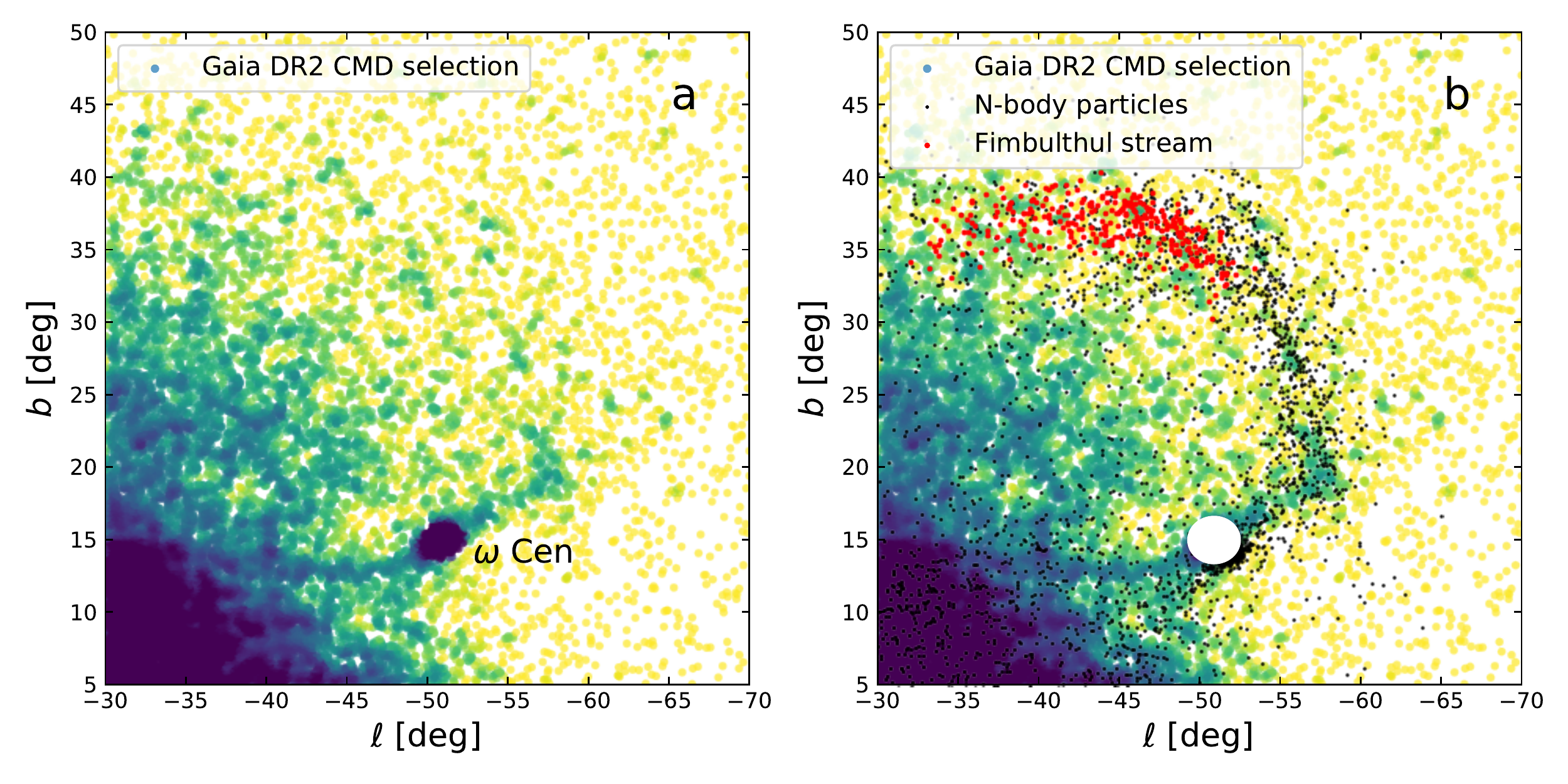}
\caption{{\bf a} Spatial distribution of stars within the CMD box of Figure~3, and that have kinematics of the $\omega$ Cen stream within $\pm 10\deg$ of the cluster as predicted by the N-body model of Figure~2. Colour encodes local stellar density (on a logarithmic scale). {\bf b} The Fimbulthul population (red) and the N-body model (black) are overlaid on the map shown in panel (a). The clear stream that is detected emanating from the cluster has similar spatial structure to the N-body model of Figure~2b.}
\end{figure}

\newpage

\section*{Methods}

\subsection{Observational Follow-up}

The six Fimbulthul stream candidate stars listed in Table~2 were observed with the ESPaDOnS spectrograph at the CFHT by observatory staff on the nights of the 24th, 25th and 28th January 2019. Two of these stars already had Gaia RVS velocity measurements, but were re-observed in order to obtain a more accurate radial velocity and to measure the chemical abundances of these stars. The targets were observed in ``star+sky'' mode, yielding very highly-resolved spectra ($R=68000$) from $369$ to $1048$~nm. The Libre-ESpRIT pipeline \cite{1997MNRAS.291..658D} was used to reduce the data, resulting in extracted and wavelength-calibrated spectra, with a normalised continuum. We measured the velocities of the target stars by cross-correlating their spectra against a spectrum of the radial velocity standard star HD182572, using the {\tt fxcor} command from the PyRAF package. Due to poor signal-to-noise, we were not able to measure the radial velocity of one of the stars (CFHT exposure number 2373599), however all others show a tight velocity grouping ($\langle v_{\rm helio} \rangle=199.7\kms$ with root mean square scatter of $5.4\kms$). This confirms the stream discovery, and suggests that the sample of Fimbulthul stream candidates listed in Table~1 is not substantially contaminated.

The metallicities of the stars were derived from the equivalent widths of the Ca~II triplet lines. The calibration we used\cite{2010A&A...513A..34S} requires knowledge of the V-band magnitude of the targets with respect to the horizontal branch. To this end we assumed that the absolute magnitude of the horizontal branch\cite{1998MNRAS.296..739G} is $V_{HB}=0.60$~mag, and used colour-transformations\cite{2010A&A...523A..48J} to convert magnitudes in the Gaia photometric bands to the V-band. For two stars (CFHT exposures 2373598 and 2373599) the spectra were of insufficient quality to allow a metallicity measurement. The other four spectra have excellent signal-to-noise, however, and result in metallicity measurements where the dominant source of uncertainty comes from the poorly-constrained distance. The metallicity values listed in Table~2 were derived assuming a mean distance to these stars of $4.1\kpc$, as predicted by the {\tt STREAMFINDER} algorithm (using the stellar populations template with a metallicity of ${\rm [Fe/H]=-1.6}$ and an age of $12.5\Gyr$). If instead we assume a common distance of $3.4\kpc$ (corresponding to the mean parallax derived above, but which as we mentioned, may be affected by contamination) the metallicities change by $+0.11$~dex. This difference provides a plausible level of uncertainty on these metallicity measurements.

\subsection{Simulation Details}

The N-body model constructed here was simulated using the NEMO software package\cite{Teuben:1995uy}. A realistic Galactic potential model\cite{Dehnen:1998tk} was used throughout this work (`model 1' of Dehnen \& Binney 1998), which consists of a disk, a thick disk, an interstellar medium, a bulge and a halo component. To perform the coordinate transformations between the models and observations, we assumed that the Sun is located $8.122\kpc$ from the Galactic centre\cite{2018A&A...615L..15G} and $17\pc$ above the Galactic mid-plane\cite{2017MNRAS.465..472K}. We take the circular velocity at the Solar neighbourhood to be $220.0\kms$, and correct for the peculiar velocity of the Sun\cite{2010MNRAS.403.1829S}. 

To model the progenitor cluster of $\omega$ Cen, we used a distribution function model that was fit to the present day structure and kinematics of the cluster, and that is also able to fit the observed strong rotation in the system. This family of rotating models\cite{2012A&A...540A..94V} has been specifically designed to describe the internal dynamics of axisymmetric systems, characterized by internal differential rotation and pressure anisotropy, and therefore it constitutes a significant improvement to the simple spherical King models. The best model has initial mass $M=(2.471\pm0.02) \times 10^6 \msun$, concentration parameter $C=1.27\pm0.01$ and core radius $R_c=127\scnp8\pm1.1$. On the plane of the sky the rotation axis of this model is inclined $12\deg$ East of North (in equatorial coordinates), with the (right-handed) north pole of the model inclined $45\deg$ towards us along the line of sight, as recently measured from Gaia DR2 proper motions\cite{2018MNRAS.481.2125B}. A total of $10^5$ particles were simulated. This model was advanced forward in time using the gyrfalcON N-body integrator\cite{Dehnen:2002vf} using a softening length for gravitational forces of $1\pc$, and a minimum time step of $2^{-18}\Gyr$.

The radial velocity and proper motion of $\omega$ Cen are now extremely well-measured\cite{2018A&A...616A..12G}, and the distance to the cluster ($d=5630\pm100\pc$) is also reasonably well constrained\cite{2018AJ....155..137B}. We first used these values to integrate an orbit backwards in time to find a first approximation to the starting position and velocity that could be used to initialize the N-body model to allow us to integrate it forwards in time to the present day. However, due to self-gravity, the dynamics of a stellar stream is non-linear, and so it can be challenging to find good initial conditions of a model that will reproduce an observed stream and remnant. Finding such initial conditions becomes harder as the total disruption time increases. The approach we took was to search for the minimum disruption time (i.e. the integration time, in increments of $0.5\Gyr$) consistent with the formation of the Fimbulthul structure. For each chosen total integration time, we surveyed many dozens of phase-space starting points in our efforts to recover a cluster with the observed present day radial velocity, proper motion and distance as $\omega$ Cen. The comparisons between the models and the observations were performed visually, attempting to find the best positional and kinematic match of the N-body stream at the sky position of $\omega$ Cen and the Fimbulthul structure. 

In our best simulation, the material that makes up the Fimbulthul stream was removed from the cluster relatively recently ($\sim 0.4\Gyr$ ago). However, we found that we require the gravitational influence of the stream particles disrupted at earlier times to reproduce the morphology and kinematics of a trailing arm resembling the Fimbulthul stream (i.e. the self-gravity of the stream appears to be important). With the N-body model used here, this required a minimum disruption time of $\sim 5\Gyr$. The best model was started ($5\Gyr$ ago) from an initial position $(x,y,z)=(3.3,-5.6,2.0)\kpc$, and with velocity $(v_x,v_y,v_z)=(98.3,20.0,-47.2)\kms$; this Galactic Cartesian coordinate system is defined so that $z$ points perpendicular to the Galactic disk with the Sun located at $(x,y,z)=(-8.122,0.0,0.017)\kpc$. In this model, the final bound structure possesses a mass of $2.03\times 10^6 \msun$, close to the initial mass of the system.  We checked for convergence by re-running the model that best fits the stream with $10^6$ particles, which showed a qualitatively identical distribution of stream stars as the $10^5$ particle simulation. We stress that this modelling effort is exploratory and is intended to show that the Fimbulthul stream is very likely the tidal stream of $\omega$ Cen; it is not a comprehensive quantitative survey of the dynamics of this system.

\subsection{Data availability}

The Gaia DR2 data on which this study was based are available at \url{https://www.cosmos.esa.int/web/gaia/dr2}. The Fimbulthul stars shown in Figure~1 are listed in Table~1.

\newpage

\newcommand{\apj}{Astrophysical Journal}
\newcommand{\apjs}{Astrophysical Journal Supplement}
\newcommand{\apjl}{Astrophysical Journal Letters}
\newcommand{\aj}{Astronomical Journal}
\newcommand{\mnras}{Monthly Notices of the Royal Astronomical Society}
\newcommand{\nat}{Nature}

\newpage

\begin{addendum}
 \item This work has made use of data from the European Space Agency (ESA) mission {\it Gaia} (\url{https://www.cosmos.esa.int/gaia}), processed by the {\it Gaia} Data Processing and Analysis Consortium (DPAC, \url{https://www.cosmos.esa.int/web/gaia/dpac/consortium}). Funding for the DPAC has been provided by national institutions, in particular the institutions participating in the {\it Gaia} Multilateral Agreement. We thank the staff of the CFHT for taking the ESPaDOnS data used here, and for their continued support throughout the project. Based on observations obtained at the Canada-France-Hawaii Telescope which is operated by the National Research Council of Canada, the Institut National des Sciences de l'Univers of the Centre National de la Recherche Scientique of France, and the University of Hawaii. This work has been published under the framework of the IdEx Unistra and benefits from a funding from the state managed by the French National Research Agency as part of the investments for the future program. PyRAF is a product of the Space Telescope Science Institute, which is operated by AURA for NASA. RAI and NFM gratefully acknowledge support from a ``Programme National Cosmologie et Galaxies'' grant. 
 \item[Competing Interests] The authors  have no
competing financial interests.
\item[Author contributions] All authors assisted in the 
development, analysis and writing of the paper. R.I., K.M. and N.M. devised the {\tt STREAMFINDER} software that detected the Fimbulthul stream. M.B. analysed the Gaia mission data to reveal the presence of the stream close to $\omega$ Cen. The initial conditions for the dynamical model of $\omega$ Cen were developed by P.B. 
\item[Correspondence] Reprints and permissions information
is available at www.nature.com/reprints. Correspondence and requests
for materials should be addressed to
R.I. (rodrigo.ibata@astro.unistra.fr).
\end{addendum}


\begin{spacing}{1.0}
\begin{center}
\footnotesize
\begin{longtable}{rrccccc}
\hline
\hline
RA J2000 & Dec J2000 & $\mu_\alpha$ & $\mu_\delta$ & $\varpi$ &  $G_0$ & $(G_{\rm BP}-G_{\rm RP})_0$ \\
${\rm [deg]}$ & ${\rm [deg]}$ & $\masyr$ & $\masyr$ & $\mas$ & ${\rm [mag]}$ & ${\rm [mag]}$ \\
\hline
 200.478245 & -25.511713 & $ -10.610 \pm 0.114$ & $  -9.710 \pm 0.112$ & $  0.391 \pm 0.073$ & 15.951 & 0.811 \\
 204.456441 & -25.702749 & $ -14.119 \pm 0.372$ & $ -10.936 \pm 0.258$ & $  0.604 \pm 0.168$ & 17.779 & 0.768 \\
 205.286684 & -23.989905 & $ -15.146 \pm 0.372$ & $ -11.223 \pm 0.274$ & $  0.223 \pm 0.156$ & 17.996 & 0.800 \\
 200.220408 & -24.769043 & $  -9.488 \pm 0.294$ & $  -9.677 \pm 0.248$ & $  0.402 \pm 0.136$ & 17.500 & 0.635 \\
 210.247730 & -24.296332 & $ -16.371 \pm 0.390$ & $ -10.787 \pm 0.264$ & $  0.274 \pm 0.171$ & 18.002 & 0.775 \\
 211.095707 & -23.212157 & $ -16.720 \pm 0.245$ & $ -11.470 \pm 0.189$ & $  0.254 \pm 0.124$ & 17.341 & 0.630 \\
 213.725755 & -23.819825 & $ -16.934 \pm 0.398$ & $ -11.491 \pm 0.310$ & $  0.081 \pm 0.192$ & 17.940 & 0.770 \\
 213.443432 & -23.784399 & $ -18.403 \pm 0.496$ & $ -10.558 \pm 0.413$ & $ -0.151 \pm 0.310$ & 18.217 & 0.744 \\
 212.585609 & -23.050072 & $ -16.477 \pm 0.516$ & $ -11.187 \pm 0.389$ & $  0.211 \pm 0.265$ & 18.411 & 0.846 \\
 206.290828 & -25.792249 & $ -15.188 \pm 0.267$ & $ -11.159 \pm 0.175$ & $  0.223 \pm 0.111$ & 17.247 & 0.666 \\
\hline
\hline
\caption{Candidate members of the Fimbulthul stellar stream. Only the first 10 lines are shown.}
\end{longtable}
\end{center}
\end{spacing}

\begin{spacing}{1.0}
\begin{center}
\scriptsize
\begin{longtable}{rrcccccccc}
\hline
\hline
RA J2000 & Dec J2000 & $G_0$ & $(G_{\rm BP}-G_{\rm RP})_0$ & $v_{\rm helio}$ & ${\rm [Fe/H]}$ & CFHT & Observation & Exposure & $S/N$ \\
${\rm [deg]}$ & ${\rm [deg]}$ & ${\rm [mag]}$ & ${\rm [mag]}$ & $[\kms]$ &              &odometer & date & [s] & ${\rm pixel^{-1}}$ \\
\hline
 200.291401 & -26.516432 & 12.167 &  1.182 & $205.52\pm0.18$ &  $-1.36\pm0.11$ & 2372977 & 2019-01-24 & 600 & 41 \\
 202.470324 & -24.439468 & 12.747 &  1.073 & $191.87\pm0.27$ &  $-1.80\pm0.11$ & 2372978 & 2019-01-24 & 600 & 34 \\
 202.470611 & -25.084004 & 13.425 &  1.018 & $202.72\pm0.28$ &  $-1.47\pm0.11$ & 2373150 & 2019-01-25 & 600 & 19 \\
 205.597014 & -25.009721 & 12.969 &  1.042 & $203.74\pm0.29$ &  $-1.58\pm0.11$ & 2373151 & 2019-01-25 & 600 & 23 \\
 208.831549 & -22.971256 & 15.323 &  0.903 & $194.74\pm2.46$ & \ldots & 2373598 & 2019-01-28 & 900 & 1.5 \\
 202.505173 & -25.937347 & 14.989 &  0.949 &   \ldots        & \ldots & 2373599 & 2019-01-28 & 600 & 1.2 \\
\hline
\hline
\caption{Fimbulthul stream candidate stars observed with CFHT/ESPaDOnS. The signal to noise ratio listed is the value per $0.006$~nm pixel at $860$~nm.}
\end{longtable}
\end{center}
\end{spacing}

\end{document}